\documentstyle[preprint,aps,epsf]{revtex}
\draft
\begin{document}
\preprint{{\bf Phys. Rev. Lett. (1999) in press.}}
\title{Differential flow in heavy-ion collisions at balance energies}
\bigskip
\author{\bf Bao-An Li\footnote{email: Bali@navajo.astate.edu}
 and Andrew T. Sustich\footnote{email: Sustich@navajo.astate.edu} }
\address{Department of Chemistry and Physics\\
Arkansas State University, P.O. Box 419\\ 
State University, AR72467-0419, USA}
\maketitle

\begin{quote}
A strong differential transverse collective flow is predicted
for the first time to occur in heavy-ion collisions at balance 
energies. We also give a novel explanation for the disappearance of 
the total transverse collective flow at the balance energies. It is 
further shown that the differential flow especially at high transverse 
momenta is a useful microscope capable of resolving the balance energy's 
dual sensitivity to both the nuclear equation of state and in-medium 
nucleon-nucleon cross sections in the reaction dynamics. \\
{\bf PACS number(s): 25.70.-z,24.10.Nz}  
\end{quote}

\newpage
One of the main goals of nuclear physics is to study properties
of the dense and hot nuclear matter created in heavy-ion collisions.
Information about the equation of state (EOS) of the hot and dense matter
is important for understanding the evolution of the early universe and
the scenario of supernova explosions. Nuclear collective flow in 
heavy-ion collisions has been found to be a useful tool for extracting 
the nuclear EOS. During the last decade, intensive theoretical and 
experimental studies have revealed considerable interesting information
about the reaction dynamics of heavy-ion collisions and the nuclear 
EOS; for a recent review see, e.g., \cite{gary99,oll98,awe98,res97}.
In particular, the excitation function of transverse collective flow 
from low to ultrarelativistic energies has been found especially 
interesting. On the high energy side, a minimum of the collective 
flow is expected in reactions crossing the phase transition region 
from hadronic matter to Quark-Gluon-Plasma\cite{oll98,hun95,ris96,li98}. 
At intermediate energies, the transverse collective flow disappears at an 
incident energy, termed the balance energy $E_{bal}$. This phenomenon 
has been well established by many experiments during the last 
decade\cite{gary99}. It has been found experimentally that the 
balance energy depends sensitively on the mass and isospin of the 
colliding nuclei as well as the impact parameter of the 
reaction\cite{wes98}. Simultaneously, much theoretical work has been devoted to understand 
the mechanism responsible for the disappearance of transverse flow 
at balance energies. It has long been suggested that at the 
balance energies the attractive scattering dominant at energies 
around 10 MeV/nucleon balances the repulsive interactions dominant 
at energies around 400 MeV/nucleon. Moreover, to extract information 
about the nuclear EOS and in-medium nucleon-nucleon cross sections, 
extensive comparisons between experimental data and theoretical 
calculations on the balance energies have been carried out, 
for a review see, e.g.,\cite{das93,lkb98}. These efforts, however, 
have been severely hindered by the dual sensitivities of the 
balance energies to both the nuclear EOS and the in-medium 
nucleon-nucleon cross 
sections\cite{mol85,bert87,tsang89,vdl92,hxu92,kla93,li93,wes93,lir96}. 
In this Letter, it is shown for the first time that there exists 
clearly a strong {\it differential} transverse collective flow at the 
balance energies. Moreover, we give a novel explanation for the disappearance 
of the transverse flow at the balance energies, and further show that 
the differential flow is a useful microscope capable of resolving the 
dual sensitivity of the balance energy. 

The anisotropic collective flow (also called directed flow) 
has been studied most commonly by analyzing the average 
transverse momentum per nucleon in the
reaction plane as a function of rapidity $y$\cite{dan85}
\begin{equation}\label{tflow}
< p_x/A >(y)=\frac{1}{A(y)}\sum_{i=1}^{A(y)} p_{ix}
=\frac{1}{dN/dy}\int p_t \frac{d^2N}{dp_tdy}<cos(\phi)>(y,p_t) dp_t, 
\end{equation}                                                                                                                                                                                                      
where $A(y)$ is the number of nucleons at rapidity $y$, $\phi$ is 
the azimuthal angle of nucleons with respect to the reaction plane and
\begin{equation}
<cos(\phi)>(y,p_t)\equiv (\frac{dN}{dp_t})^{-1}
\int cos(\phi)\frac{d^2N}{dp_td\phi}d\phi.
\end{equation}
A nonvanishing $<cos(\phi)>(y,p_t)$ indicates the existence of an 
azimuthally anisotropic transverse flow at the rapidity $y$ and 
transverse momentum $p_t$, we name it the {\it differential} flow. 
In particular, adopting the Fourier expansion\cite{vol97,pos98} 
\begin{equation}
\frac{d^2N}{dp_td\phi}=\frac{dN}{dp_t}
[1+\sum_{i=1}^{i=\infty}2v_i(y,p_t)cos(i\phi)],
\end{equation}
one finds that $<cos(\phi)>(y,p_t)=v_1(y,p_t)$ is the strength 
of the azimuthal angle distribution to the first order. 
Information about the nuclear EOS and in-medium cross sections can
be preserved and revealed more completely by decoupling the 
differential flow from the integrand in Eq.\ \ref{tflow}. The analysis 
of $v_1(y,p_t)$ in heavy-ion collisions at high energies has been 
found especially useful for studying properties of the transverse 
collective flow and the nuclear EOS\cite{vol97,pos98,zha95,lik96}. 
At the balance energies the total in-plane average transverse momentum 
of Eq.\ \ref{tflow} vanishes in almost the whole range of rapidity. 
It is thus sufficient to study the $p_t$ dependence of differential
flow around the projectile and/or target rapidities where transverse 
flow peaks should it exist at all. 

\begin{figure}[htp]
\setlength{\epsfxsize=14truecm}
\centerline{\epsffile{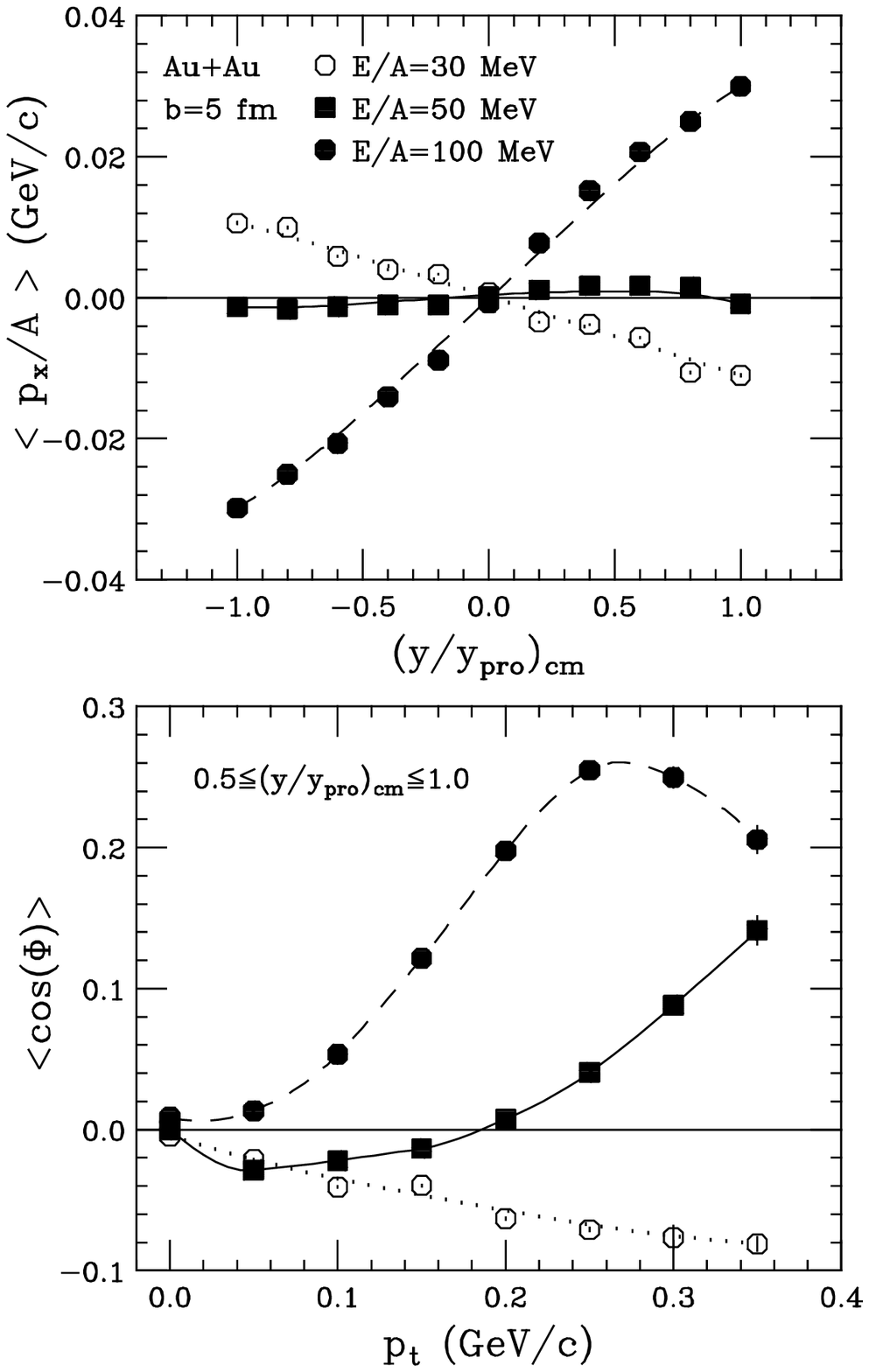}}
\caption{Total (upper) and differential (lower) transverse flow analysis
for the reaction of Au+Au at an impact parameter of 5 fm and
beam energies of 30, 50 and 100 MeV/nucleon.}
\label{fig1}
\end{figure}

Much of the current understanding about the nuclear 
EOS\cite{pan93,zha94} and in-medium nucleon-nucleon cross
sections\cite{hxu92,kla93,li93,wes93} comes from comparing 
flow measurements with predictions based on transport theories 
such as the Boltzmann-Uehling-Uhlenbeck (BUU) model. 
Details of the BUU model used in the present study can be found 
in refs. \cite{libauer1,libauer2,lik97}. To identify the balance 
energy we first performed the standard transverse momentum analysis 
for the reaction of Au+Au at an impact parameter of 5 $fm$ and 
several beam energies. Typical results of this analysis 
are shown in the upper window of Fig.\ \ref{fig1}. These calculations 
were carried out by using the stiff nuclear EOS of compressibility 
$K=380$ MeV and the free-space nucleon-nucleon cross sections given 
in refs. \cite{libauer1,libauer2,lik97}. It is seen that the 
transverse flow is attractive, almost zero and repulsive 
at the beam energy of 30, 50 and 100 MeV/nucleon, respectively.
To investigate the differential flow, the value of $<cos(\phi)>$ 
is shown as a function of $p_t$ in the lower window for nucleons 
with rapidities in the range of $0.5\leq (y/y_{proj})_{cms}\leq 1.0$. 
Although quantitatively different, our results are qualitatively 
independent of the chosen rapidity bin and impact parameters\cite{ls99}. 
Of course, $<cos(\phi)>(p_t)$ reverses its sign on the negative rapidity
side. The reactions at beam energies of 30 and 100 MeV/nucleon show clearly
the characteristic attractive and repulsive differential flow, 
respectively, as one expects in the whole range of $p_t$. It is most 
interesting to see clearly also a strong differential flow at the 
balance energy of 50 MeV/nucleon in the whole range of $p_t$. It 
changes from being attractive to repulsive at a balance transverse 
momentum of about $p_{bal}=0.2 GeV/c$. Although altogether having a 
zero average transverse momentum in the reaction plane particles 
with higher and lower $p_t$ move preferentially towards the positive 
and negative flow directions, respectively. This is because the lower
$p_t$ particles are more affected by the attractive mean field while
the higher $p_t$ particles are more affected by the 
repulsive nucleon-nucleon scatterings. It is thus clear that the disappearance 
of transverse flow is due to the cancelling of positive and negative 
differential flow of particles with high and low transverse momenta. 
We also notice that the magnitude of the observed differential flow 
at the balance energy is compatible with the azimuthal asymmetry 
normally observed in heavy-ion collisions at intermediate 
energies\cite{wes10}. A comparison of the two analyses above clearly 
indicates that the differential flow probes more microscopically the 
interesting features which are otherwise smeared out by the 
integration in the standard flow analysis. 
   
\begin{figure}[htp]
\setlength{\epsfxsize=14truecm}
\centerline{\epsffile{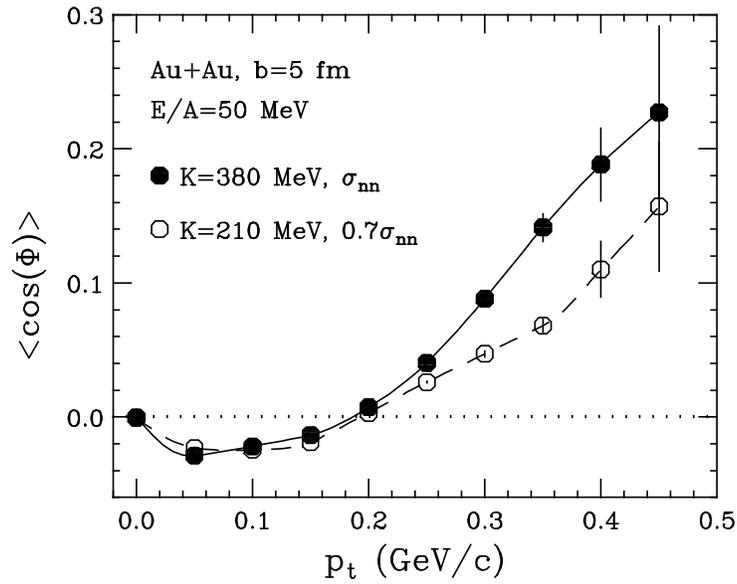}}
\caption{Differential flow analysis for Au+Au at an impact parameter 
of 5 fm using two different parameter sets leading to the 
same balance energy of 50 MeV/nucleon.}
\label{fig2}
\end{figure}

It is well known that the transverse collective flow in general, 
and the balance energy in particular are dually sensitive to both 
the nuclear EOS and the in-medium nucleon-nucleon cross sections. 
For the Au+Au reaction we found that a soft EOS of compressibility 
$K=210$ MeV with a $30\%$ reduction in the nucleon-nucleon cross 
sections results in about the same balance energy as the stiff 
one with the free-space nucleon-nucleon cross sections. This is qualitatively  
in agreement with the previous findings\cite{hxu92,kla93,li93,wes93}. 
More quantitatively, an average in-plane transverse momentum per 
nucleon of $0.86$ and $-0.02$ MeV/c in the range of $0.5\leq (y/y_{proj})_{cms}\leq 1.0$ 
is obtained by using of the stiff and soft EOS, respectively. 
Both of these values are equivalent to zero transverse flow 
within the statistical errors of the calculations. 
Shown in Fig.\ \ref{fig2} is an analysis of the differential flow 
for the above two cases which give the same balance energy of 
50 MeV/nucleon. A clear, experimentally distinguishable separation 
in the differential flow is seen at high transverse momenta above 
about 0.2 GeV/c. The soft EOS with the reduced in-medium cross sections 
lowers the differential flow at high transverse momenta by about a 
factor of 2. A counter balancing shift in the $p_t$ spectrum 
$<p_tdN/dp_t/A>$ is found simultaneously and will be published 
elsewhere\cite{ls99}. The differential flow is thus a useful 
microscope capable of resolving the balance energy's dual sensitivity 
to both the nuclear EOS and the in-medium cross sections.

\begin{figure}[htp]
\setlength{\epsfxsize=14truecm}
\centerline{\epsffile{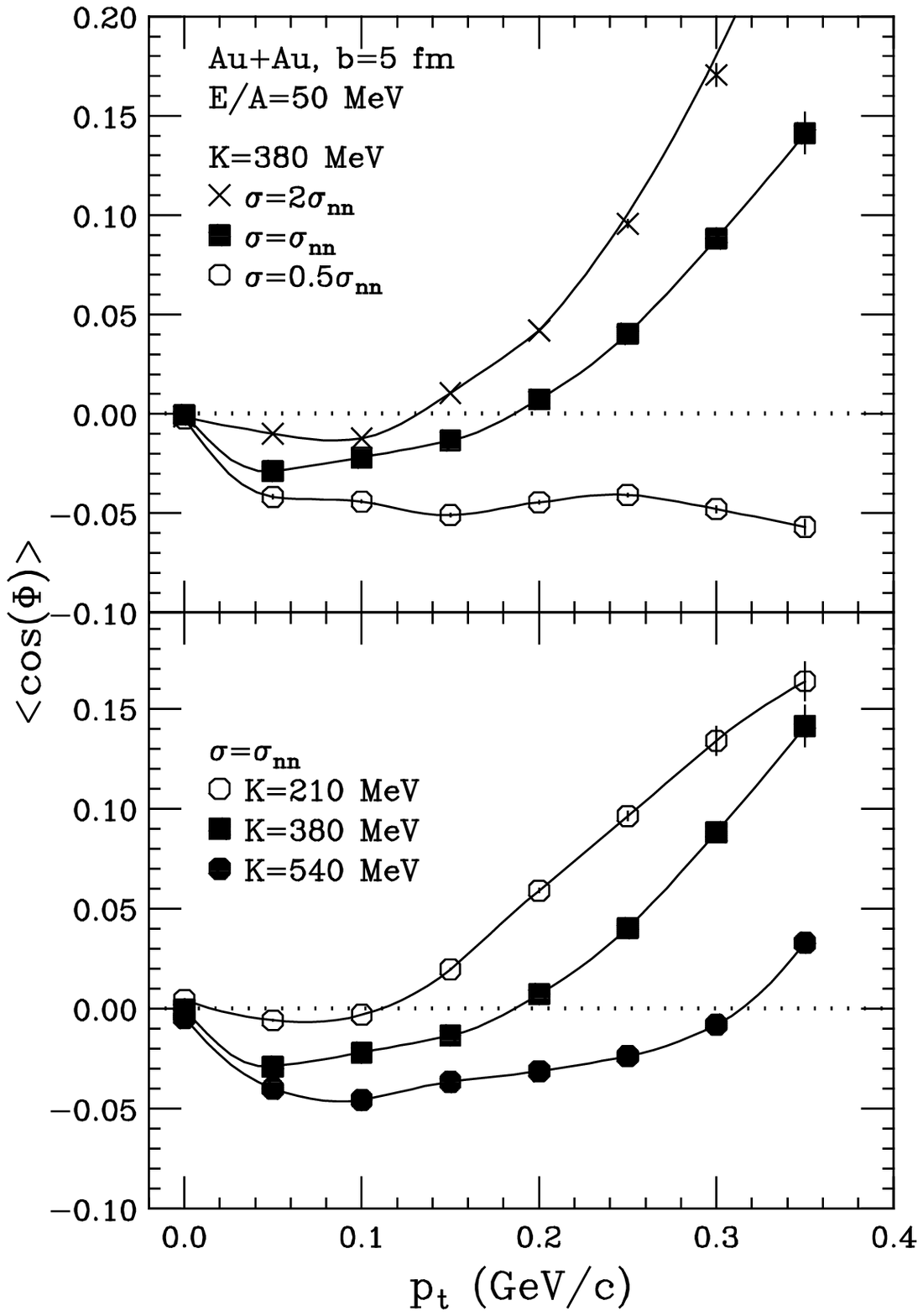}}
\caption{Dependence of the differential flow on the in-medium cross section
(upper) and compressibility (lower) for Au+Au at an impact parameter 
of 5 fm and a beam energy of 50 MeV/nucleon.}
\label{fig3}
\end{figure}

Our findings here can be further understood by studying the competitive 
roles of the nuclear EOS and in-medium nucleon-nucleon cross sections 
in forming the differential flow at the balance energies. Results of 
such a study is shown in Fig.\ \ref{fig3} for the Au+Au reaction at
$E/A=50$ MeV/nucleon. In the upper window, the differential flow is 
studied by varying the in-medium nucleon-nucleon cross section 
$\sigma$ in the range of current theoretical predictions\cite{gqli,alm1,alm2}. 
It is seen that the differential flow increases as the $\sigma$ 
increases because it enhances the repulsive scatterings. In the lower 
window, the effect of the nuclear mean field is studied by varying the  
compressibility $K$ in the range currently considered in the 
literature of astro- and nuclear physics\cite{baron85,ter87,hart99}. 
In the energy range studied here the mean field is attractive and its
effect on the differential flow increases with the increasing compressibility.
The differential flow at balance energies is a remnant of the
competition between the in-medium cross section and the nuclear mean 
field. Since a $30\%$ reduction in the $\sigma$ reduces the 
differential flow more than the increase caused by a softening of the 
nuclear EOS from $K=380$ to $210$ MeV, especially at high transverse 
momenta, our findings in Fig.\ \ref{fig2} are easily understandable. 
The differential flow reveals more microscopically and directly the 
competition between the negative scattering due to the attractive 
mean field and the positive scattering due to the repulsive 
nucleon-nucleon collisions. This strong competition also makes the 
differential flow a much more sensitive probe than the balance energy 
itself. It is therefore interesting to note that the balance energy 
has been investigated extensively as a function of mass, isospin and 
impact parameter of the reaction in many experiments during the last 
decade. The differential flow analysis can thus be carried out using 
available data. This study will be both physically very interesting  
and highly economical since the data already currently exists. 
In addition, our present work 
has important physical implications to the study of ``zero'' flow 
of secondary particles, such as, pions and kaons, produced 
in heavy-ion collisions. Many interesting predictions relating their 
collective flow and in-medium dispersion relations have been 
made\cite{kol96}, however, an almost ``zero'' flow have been found 
experimentally\cite{rei98,ogi98}. Our differential flow analysis for
these particles is in progress and will be published elsewhere.   
   
In summary, it is predicted for the first time that a strong 
differential collective flow exists at balance energies where the
normal transverse collective flow disappears. We also give a novel 
explanation for the disappearance of total transverse flow at the 
balance energies. It is further shown that the differential flow 
especially at high transverse momenta is a useful microscope capable 
of resolving the balance energy's dual sensitivity to both the 
nuclear EOS and the in-medium nucleon-nucleon cross sections in the 
reaction dynamics. Experimental analysis of differential flow at 
balance energies using available data will be very interesting and 
fruitful.

We would like to thank Dr. C.M. Ko and Dr. Bin Zhang for their 
critical reading of the manuscript.

\newpage

\newpage

\newpage


\begin{thebibliography}{99}

\bibitem{gary99}G.D. Westfall and J. P\'eter, 
Ann. Rev. Nucl. Part. Sci. (1999) to be published.

\bibitem{oll98}J.-Y. Ollitrault, Quark Matter'97, 
Nucl. Phys. {\bf A638}, 195c (1998). 

\bibitem{awe98}T.C. Awes, Nucl. Phys. {\bf A630}, 499c (1998).

\bibitem{res97}W. Reisdorf and H.G. Ritter, 
Ann. Rev. Nucl. Part. Sci. {\bf 47}, 663 (1997).

\bibitem{hun95}C.M. Hung and E. V. Shuryak, 
Phys. Rev. Lett. {\bf 75}, 4003 (1995).

\bibitem{ris96}D.H. Rischke and M. Gyulassy, 
Nucl. Phys. {\bf A597}, 701 (1996).  

\bibitem{li98}B.A. Li and C.M. Ko, Phys. Rev. C{58}, R1382 (1998).

\bibitem{wes98}G.D. Westfall, Nucl. Phys. {\bf A630}, 27c (1998).

\bibitem{das93}S. Das Gupta and G.D. Westfall, 
	Physics Today, {\bf 46}(5), 34 (1993).

\bibitem{lkb98}B.A. Li, C.M. Ko and W. Bauer, 
Int. Jou. of Mod. Phys. E{\bf 7}, 147 (1998).

\bibitem{mol85}J. Molitoris and H. St\"ocker, 
		Phys.\ Lett.\ {\bf B162}, 47 (1985).

\bibitem{bert87}G.F. Bertsch, W.G. Lynch and M.B. Tsang, 
	Phys.\ Lett.\ {\bf B189}, 738 (1987).

\bibitem{tsang89}M.B. Tsang, G.F. Bertsch, W.G. Lynch and M. Tohyama,
	Phys.\ Rev. C{\bf 40}, 1685 (1989).

\bibitem{vdl92}V. de la Mota, F. Sebille, M. Farine, B. Remaud and P. Schuck, 
	Phys.\ Rev.\ C{\bf 46}, 677 (1992).

\bibitem{hxu92}H.M. Xu, Phys.\ Rev.\ Lett.\ {\bf 67}, 2769 (1992); 
	Phys.\ Rev.\ C{\bf 46}, R392 (1992).

\bibitem{kla93}D. Klakow, G. Welke and W. Bauer, 
Phys.\ Rev. C{\bf 48}, 1982 (1993).

\bibitem{li93}B.A. Li, Phys.\ Rev.\ C{\bf 48}, 2415 (1993).

\bibitem{wes93} G.D. Westfall, W. Bauer {\it et al.}, Phys.\ Rev.\ Lett.\, 
      {\bf 71}, 1986 (1993).

\bibitem{lir96}B.A. Li, Z.Z. Ren, C.M. Ko and S.J. Yennello,
Phys. Rev. Lett. {\bf 76}, 4492 (1996).

\bibitem{dan85}P. Danielewicz and G. Odyniec, Phys. Lett. {\bf B157}, 146 (1985).

\bibitem{vol97}S.A. Voloshin, Phys. Rev. C{\bf 55}, R1630 (1997).

\bibitem{pos98}A.M. Poskanzer and S.A. Voloshin, 
Phys. Rev. C{\bf 58}, 1671 (1998).

\bibitem{zha95}Y. Zhang and J.P. Wessels for E877 collaboration,
Nucl. Phys. {\bf A590}, 557c (1995).

\bibitem{lik96}B.A. Li, C.M. Ko and G. Q. Li, 
Phys. Rev. C{\bf 54}, 844 (1996).

\bibitem{pan93}Q. Pan and P. Danielewicz, 
Phys.\ Rev.\ Lett.\ {\bf 70}, 2062 (1993).

\bibitem{zha94}J. Zhang, S. Das Gupta and C. Gale, 
	Phys.\ Rev.\ C{\bf 50}, 1617 (1994).

\bibitem{libauer1} B.A. Li and W. Bauer, 
      Phys. Rev. C{\bf 44}, 450 (1991). 

\bibitem{libauer2} B.A. Li, W. Bauer and G.F. Bertsch, 
      Phys. Rev. C{\bf 44}, 2095 (1991). 

\bibitem{lik97}B.A. Li, C.M. Ko and Z.Z. Ren, 
Phys. Rev. Lett. {\bf 78}, 1644 (1997).

\bibitem{ls99}B.A. Li and A. T. Sustich, (1999) to be published.

\bibitem{wes10}G.D. Westfall for the $4\pi$ collaboration,
In Advances in Nuclear Dynamics, Eds. J. Harris, A. Mignerey 
and W. Bauer, p. 274, World Scientific (Singapore), 1994. 

\bibitem{gqli}G.Q. Li and R. Machleidt, Phys. Rev. C{\bf 48}, 1702 (1993);
{\it ibid} C{\bf 49}, 566 (1994).

\bibitem{alm1}T. Alm, G. R\"opke and M. Schmidt, Phys. Rev. C{\bf 50}, 31 (1994).

\bibitem{alm2}T. Alm, G. R\"opke, W. Bauer, F. Daffin and M. Schmidt, 
Nucl. Phys. {\bf A587}, 815 (1995).

\bibitem{baron85}E. Baron et al., Phys. Rev. Lett. {\bf 55}, 126 (1985);
Nucl. Phys. {\bf A440}, 744 (1985).

\bibitem{ter87}B. Ter Haar and R. Malfliet, Phys. Lett. {\bf B172}, 10 (1986); 
Phys. Rep. {\bf 149}, 207 (1987).

\bibitem{hart99}Ch. Hartnack et al., preprint Nucl-th/9901087. 

\bibitem{kol96}C.M. Ko and G.Q. Li, J. Phys. G{\bf 22}, 1673 (1996).

\bibitem{rei98}W. Reisdorf, Nucl. Phys. {\bf A630}, 15c (1998).

\bibitem{ogi98}C.A. Ogilvie, Nucl. Phys. {\bf A630}, 571c (1998);
in Proc. of Strange Matter'98, to be published.
\end{thebibliography}
\end{document}